\documentclass[conference]{IEEEtran}

\IEEEoverridecommandlockouts

\usepackage{hyperref}
\usepackage{amsmath,amssymb,amsfonts}
\usepackage{algorithmic}
\usepackage{graphicx}
\usepackage{textcomp}
\usepackage{xcolor}
\usepackage{siunitx}

\def\BibTeX{{\rm B\kern-.05em{\sc i\kern-.025em b}\kern-.08em
    T\kern-.1667em\lower.7ex\hbox{E}\kern-.125emX}}

\begin{document}

\title{Towards Quantifying Two-Mode Correlation Linewidths in Quantum Circuits\\

\thanks{This work is partially supported by the European project MiSS. 
MiSS is funded by the European Union through the Horizon Europe 2021-2027 Framework Programme, Grant agreement ID: 101135868.
This work is partially supported by the Italian project CalQuStates. The PRIN 2022 CalQuStates project received funding by the European Union – Next Generation EU.
This work is partially supported by the Italian project Quantum Radar. The Quantum Radar project (CUP ZAE31B3A7C) received funding by the Italian Ministry of Defence.
This publication has been partially funded by the Italian Ministry of University and Research (MUR) in the framework of the continuing nature project “NEXT GENERATION METROLOGY”, under the allocation of the Ordinary Fund for research institutions (FOE) 2023 (Ministry Decree n. 789/2023).
}
}

\author{\IEEEauthorblockN{1\textsuperscript{st} Alessandro Alocco}
\IEEEauthorblockA{\small \textit{Politecnico di Torino} \\
Torino, Italy\\
+39 011 3919 434\\
alessandro.alocco@polito.it}
\and
\IEEEauthorblockN{2\textsuperscript{nd} Andrea Celotto}
\IEEEauthorblockA{\small \textit{Politecnico di Torino} \\
Torino, Italy\\
+39 011 3919 434\\
andrea.celotto@polito.it}
\and
\IEEEauthorblockN{3\textsuperscript{rd} Luca Fasolo}
\IEEEauthorblockA{\small \textit{INRiM} \\
Torino, Italy \\
+39 011 3919 435\\
l.fasolo@inrim.it}
\and
\IEEEauthorblockN{4\textsuperscript{th} Bernardo Galvano}
\IEEEauthorblockA{\small \textit{Università di Palermo} \\
Palermo, Italy \\
+39 011 3919 434\\
bernardo.galvano@you.unipa.it}
\and
\IEEEauthorblockN{5\textsuperscript{th} Emanuele Palumbo}
\IEEEauthorblockA{\small \textit{Politecnico di Torino} \\
Torino, Italy\\
+39 011 3919 434\\
emanuele.palumbo@polito.it}
\and 
\IEEEauthorblockN{6\textsuperscript{th} Luca Oberto}
\IEEEauthorblockA{\small \textit{INRiM} \\
Torino, Italy \\
+011 3919 327\\
l.oberto@inrim.it}
\and
\IEEEauthorblockN{7\textsuperscript{th} Luca Callegaro}
\IEEEauthorblockA{\small \textit{INRiM} \\
Torino, Italy \\
+39 011 3919 435\\
l.callegaro@inrim.it}
\and \quad
\IEEEauthorblockN{8\textsuperscript{th} Felice Francesco Tafuri}
\IEEEauthorblockA{\small \textit{Keysight Technologies Italy S.r.l.} \\
Rome, Italy \\
+34 933434707\\
francesco.tafuri@keysight.com}
\and
\IEEEauthorblockN{9\textsuperscript{th} Patrizia Livreri}
\IEEEauthorblockA{\small \textit{Università di Palermo} \\
Palermo, Italy \\
+39 091 238 60249\\
patrizia.livreri@unipa.it}
\and
\IEEEauthorblockN{10\textsuperscript{th} Emanuele Enrico}
\IEEEauthorblockA{\small \textit{INRiM} \\
Torino, Italy \\
+39 011 3919 433\\
e.enrico@inrim.it}
}
\maketitle

\begin{abstract}
This paper aims to quantify the linewidth of two-mode correlations in Traveling Wave Parametric Amplifiers (TWPAs). 
Artifacts induced by data acquisition and processing, such as windowing effects and acquisition time, are examined to understand their influence on the linewidth estimation of these correlations. The findings underscore the significance of acquisition parameters in optimizing two-mode correlation measurements, enhancing device characterization for quantum applications.
\end{abstract}


\section{Introduction}

\begin{figure}[htbp]
    \centering
\includegraphics[width=0.9\linewidth]{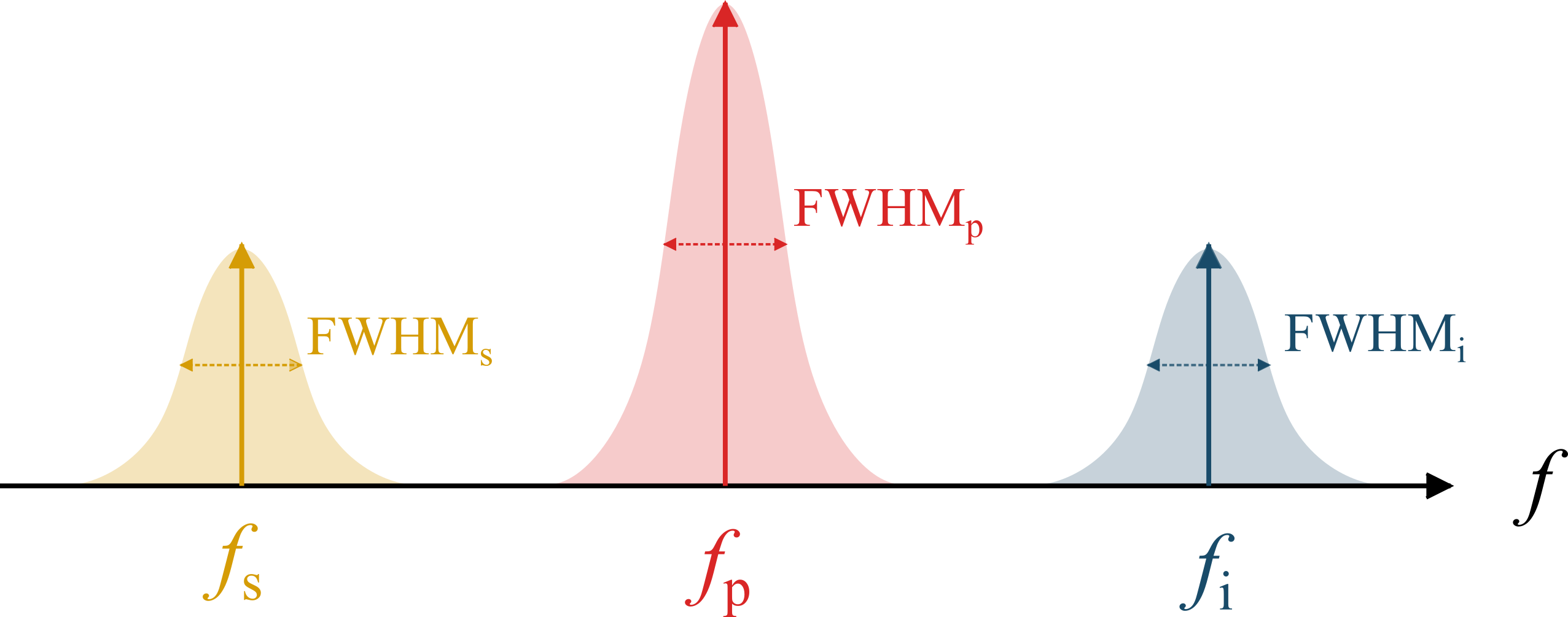}
    \caption{Spectral representation of a four-wave mixing process. The signal and idler frequencies $f_\text{s}$ and $f_\text{i}$ are symmetric with respect to the pump frequency $f_\text{p}$. Each tone has its own linewidth, estimated by the Full Width at Half Maximum (FWHM).}
    \label{spectrum}
\end{figure}

Recent advancements in superconducting nonlinear circuits, particularly Traveling Wave Parametric Amplifiers (TWPAs), have revolutionized the ability to generate and amplify quantum states in the microwave regime \cite{Colloquium_Nonlinear_metamaterials}. These devices owe their effectiveness to their broadband operation and tunable nonlinear properties \cite{Jung_2014}. At the heart of TWPAs lies the wave-mixing process, which is critical for parametric amplification, entanglement generation, and squeezing \cite{Nonlinear_Optics_Third_Edition}.
Two-mode squeezing stands out for its pivotal role in quantum sensing \cite{Quantum_sensing}, continuous-variable quantum computing \cite{Quantum_information_with_continuous_variables}, and quantum illumination \cite{Detecting}. This phenomenon consists of the generation of entangled electromagnetic modes, which are valuable resources for quantum protocols \cite{Quantum_information_with_continuous_variables}. Maximum correlation arises when energy conservation is fulfilled during the wave-mixing process between a pump tone with frequency $f_\text{p}$ and two other modes, referred to as signal and idler with respective frequencies $f_\text{s}$ and $f_\text{i}$ \cite{Eichler}.
For a four-wave mixing (4WM) process in a TWPA, the energy conservation relation is 
\begin{equation}
\label{freqmatch}
   \Delta f \equiv 2f_\text{p} - f_\text{s} - f_\text{i} = 0 ,
\end{equation}
which means that the signal and idler $f_\text{s}$ and $f_\text{i}$ are symmetric in the frequecy domain with respect to to the pump frequency $f_\text{p}$ (see Fig. \ref{spectrum}). This study aims to investigate the linewidth of the 4WM phenomenon using two-mode correlation and to identify the factors that influence it.
Both intrinsic factors and artifacts induced by data acquisition can broaden the linewidth and cause frequency shifts. The latter arise from factors such as windowing effects and acquisition time. 
In contrast, intrinsic factors are related to physical properties, such as imperfections in the TWPA and frequency dispersion in the signal and idler paths.
Optimizing the data acquisition and processing is then beneficial for gaining new insights into TWPA characterization and their application in quantum illumination \cite{fasolo2021josephsontravelingwaveparametric}.

\section{Two-Mode Correlations in TWPAs}
\label{Theoretical Framework}
\begin{figure}[htbp]
    \centering
    \includegraphics[width=0.9\linewidth]{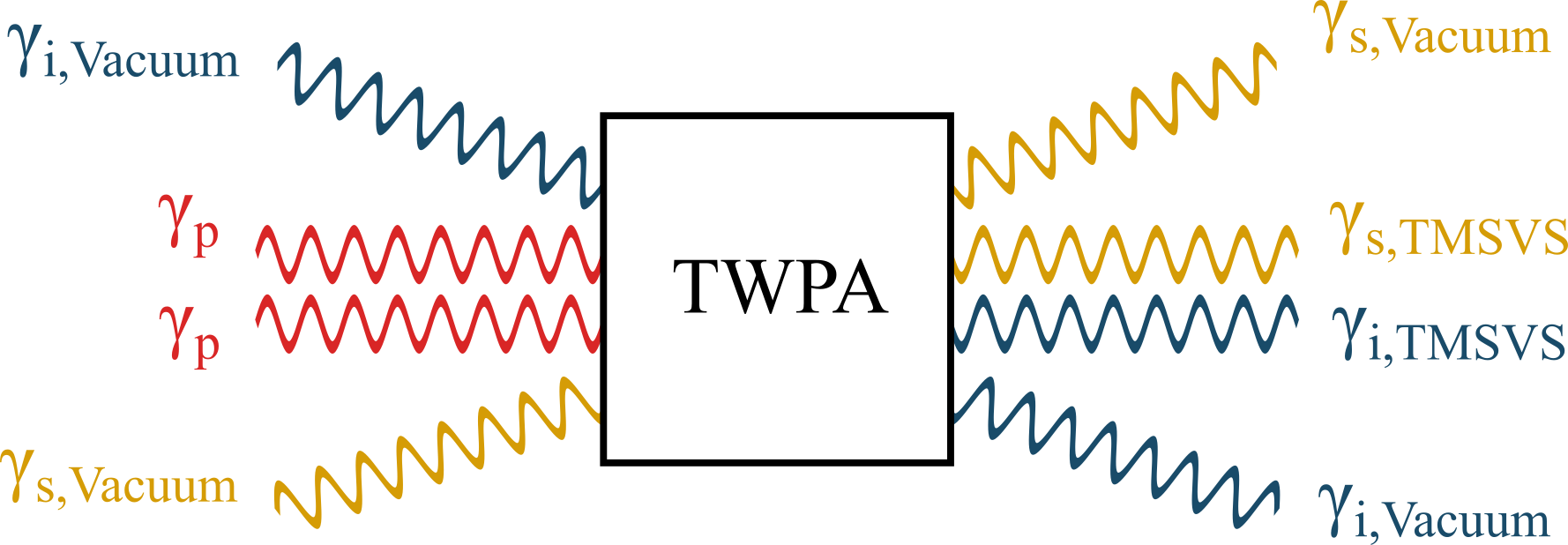}
    \caption{Pictorial representation of the generation of the TMSVS in a 4WM process inside a TWPA. Two pump photons $\gamma_{\text{p}}$ interact with a signal $\gamma_{\text{s,Vacuum}}$ and an idler $\gamma_{\text{i,Vacuum}}$ photon inside the TWPA. As a result of this process the two pump photons are converted into entangled pair of signal $\gamma_{\text{s,TMSVS}}$ and idler $\gamma_{\text{i,TMSVS}}$ photons.
    }
    \label{fig_4WM}
\end{figure}

Two-mode correlations in a TWPA arise from a parametric process driven by a pump tone, which generates correlated photon pairs, known as the signal and idler.
In the case of a quantum-limited TWPA, the use of a vacuum state as the input enables the generation of an entangled pair of signal and idler photons: the Two-Mode Squeezed Vacuum State (TMSVS) \cite{Eichler} \cite{Esposito_2022}.
The process is described in Fig. \ref{fig_4WM}. 
The interaction Hamiltonian of a quantum-limited TWPA, operating in the non-degenerate regime (i.e. with $f_\text{s} \neq f_\text{i}$) can be expressed as \cite{article1}:
\begin{equation}
\label{hamiltonian}
\hat{H}_{\text{int}} = i \hbar k \left[ e^{i\theta} \hat{a}_\text{s}^\dagger \hat{a}_\text{i}^\dagger + e^{-i\theta} \hat{a}_\text{s} \hat{a}_\text{i}  \right],
\end{equation}
where $\hat{a}_\text{s}$ ($\hat{a}_\text{s}^\dagger$) and $\hat{a}_\text{i}$ ($\hat{a}_\text{i}^\dagger$) are the annihilation (creation) operators of the signal and idler modes, respectively. The parameter $k$ represents the nonlinear coupling, given by the susceptibility of the TWPA and by the pump field amplitude. The phase $\theta$ is related to the mismatch between pump, signal and idler phases ($\phi_\text{p}$, $\phi_\text{s}$, and $\phi_\text{i}$) along the TWPA.
The wave-mixing is favored when $\theta = 0$. The particular expression for the phase mismatch depends on the wave-mixing process involved in the interaction.
For a 4WM process, the phase-matching condition is $\theta \equiv 2\phi_\text{p}-\phi_\text{s}-\phi_\text{i} = 0$. This condition maximizes the nonlinear interaction, ensuring two-mode correlations between the signal and idler mode quadratures (IQ data), which in this paper will be referred to as X (in-phase) and P (in-quadrature) quadratures, respectively.
Starting from $\hat{a}$ and $\hat{a}^\dagger$, the in-phase quadrature for signal and idler are defined as 
\begin{equation}
    \hat{X}_\text{s,i}^\theta = \frac{1}{2}(\hat{a}_\text{s,i} e^{-i\theta} + \hat{a}_\text{s,i}^\dagger e^{i\theta}).
\end{equation}
The TWPA's input-output relation for ${\hat{X}}_\text{s,i}$ is derived from Eq. \ref{hamiltonian}:
\begin{equation}
\label{xout}
    \hat{X}_{\text{s,i}}^{\text{out}} = \sqrt{G_\text{s,i}}\,\hat{X}_{\text{s,i}}^{\text{in,0}}  + \sqrt{G_\text{i,s}-1}\,\hat{X}_{\text{i,s}}^{\text{in,}\theta}, \\
    \\
\end{equation}
where $G_\text{s}$ ($G_\text{i}$) is the TWPA gain at signal (idler) frequency. The quadratures of the input vacuum states for signal and idler frequencies, $\hat{X}_{\text{s,i}}^{\text{in}}$, satisfy the uncertainty condition 
$\sigma^2(X_{\text{s,i}}^{\text{in}}) = 1/4$
\cite{PhysRevA.79.063826}.
Under vanishing detuning ($\Delta f = 0$), ideal phase matching ($\theta = 0$), and equal gain at signal and idler frequencies ($G_{s}=G_{i}=G$), the variance of the collective quadrature $\hat{X} = {\hat{X}}_\text{s}^{\text{out}} - {\hat{X}}_\text{i}^{\text{out}}$ is minimized:
\begin{equation}
        \sigma^2(X) = \frac{1}{2}e^{-2 { \text{ arcosh}
}(\sqrt{G})}.\\
        \\
\end{equation}
This condition corresponds to maximal squeezing, where the squeezing is defined as $S = 10 \log_{10} \left(\sigma^2(X) / 0.5 \right)$ \cite{Esposito_2022}.
Since in general $G_\text{s} \neq G_\text{i}$, $\sigma^2(X)$ is not a fair estimator of squeezing correlations. Instead, the Pearson correlation coefficient 
\begin{equation}
    \rho(X_\text{s}, X_\text{i}) = {\text{cov}(X_\text s,X_\text i)}/({\sigma(X_\text s)\sigma({X_\text i}))}  
\end{equation}
between the signal and idler observed quadratures $X_\text{s,i}^{\text{}}$ is considered, since asymmetries in the gain profile are intrinsically taken into account.
This makes it a more reliable estimator for two-mode correlations, even if it does not express any quantum property of the phenomenon.
\section{Two-Mode Correlation Experiment}
\begin{figure}[htbp]
    \centering    \includegraphics[width=0.9\linewidth]{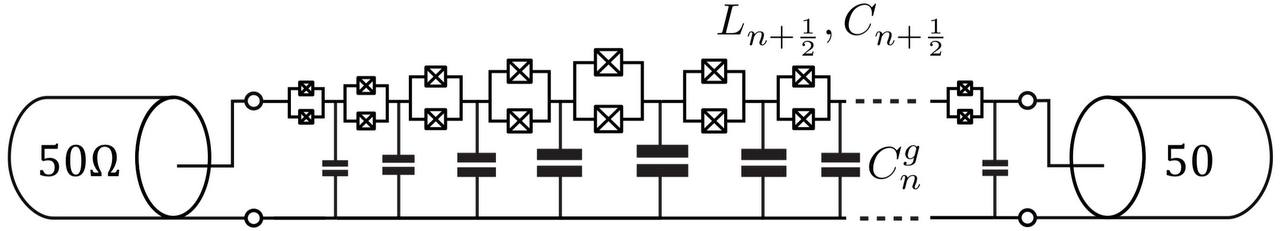}
    \caption{Electrical sketch of the TWPA by SilentWaves. The device consists of a $50\text{ }\Omega$ impedance-matched nonlinear transmission line based on Josephson junctions (crossed box symbol). The phase-matching is realized by engineering the
    dispersion relation of the transmission line using
    a spatial modulation of its impedance. To do so, the electrical parameters, including the Josephson
inductance $L_{n+{1/2}}$ and capacitance $C_{n+1/2}$, as well as the capacitance
between each island and the ground plane $C^g_{n}$, are periodically modulated. Reproduced from Fig. 1 in \cite{argo}.}
    \label{fig:argo}
\end{figure}

The experiment consists in generating two-mode correlated radiation and acquiring the IQ data of the signal and idler.
Two-mode squeezed radiation is generated by a pump tone injected into the TWPA that gets coupled with the vacuum state to produce a two-mode squeezed vacuum state, as described in Section \ref{Theoretical Framework}. The two-mode correlated radiation is then digitized and IQ-demodulated for further analysis.
The device under test in the following experiments is a commercial TWPA developed by Silent Waves, operating in the 4WM regime \cite{argo}.
Its electrical sketch is shown in Fig. \ref{fig:argo}.

\subsection{The Experimental Setup}
\label{Exp setup}
\begin{figure}[htbp]
\centerline{\includegraphics[width=\linewidth]{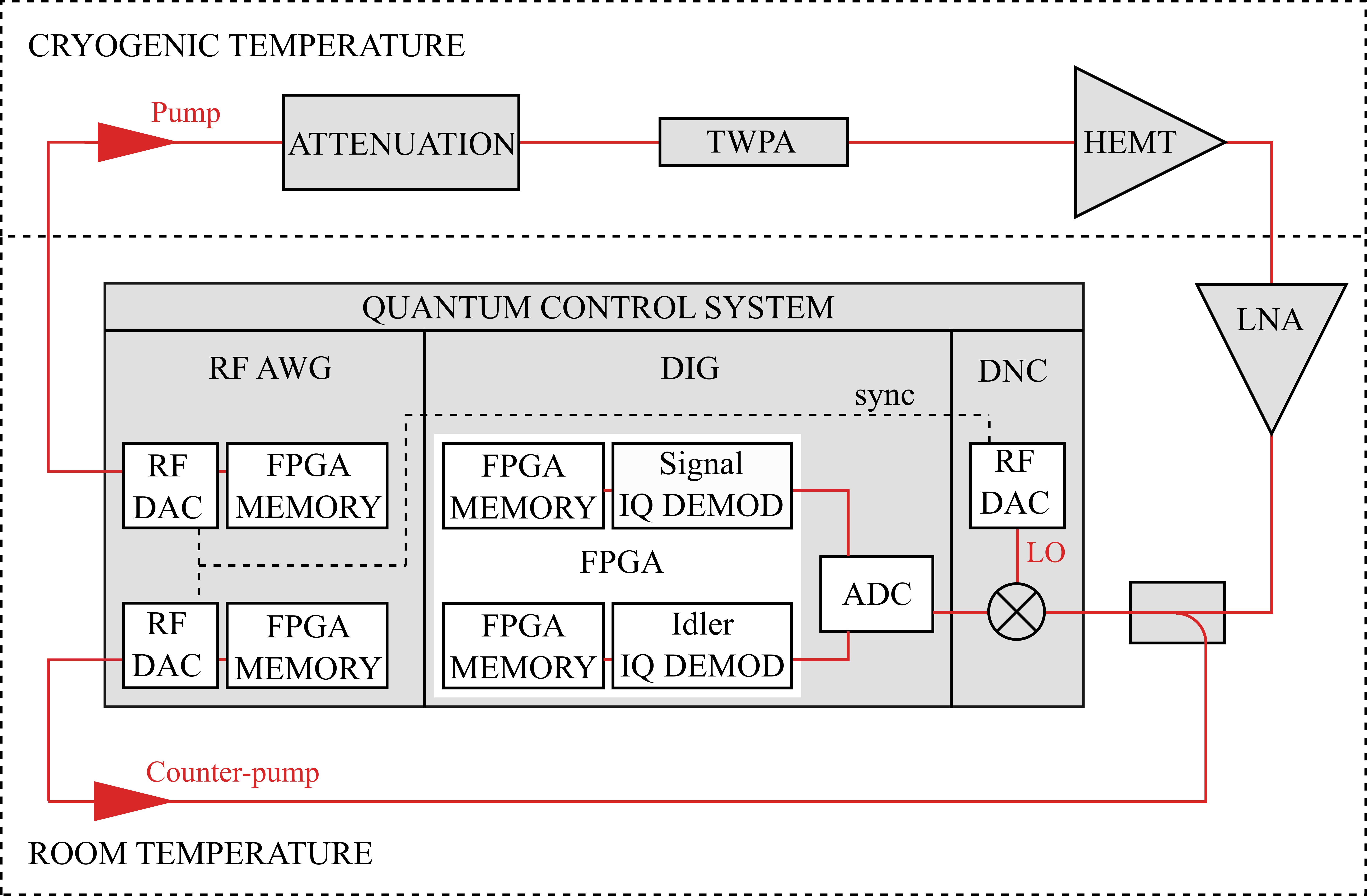}}
\caption{
Schematic representation of the cryogenic and room temperature setup. See decription in Sec. \ref{Exp setup}.
}
\label{fig_Experimental setup.}
\end{figure}

The experimental setup is shown in Fig. \ref{fig_Experimental setup.}.
An Arbitrary Radio Frequency Arbitrary Waveform Generator
(RF AWG) is used to feed the TWPA with a pump pulse generated by a Radio Frequency Digital to Analog Converter (RF DAC). This pulse
enters the input line of a Dilution Refrigerator (DR)
having a base temperature of about 20 mK. Before reaching the TWPA, it is
properly attenuated (about 60 dB in total) to wash out
thermal fluctuations. The TWPA output passes a 60 dB
isolator stage located at 20 mK (not shown in the picture) and is then amplified by a High-Electron-Mobility Transistor (HEMT) low noise amplifier located at the 3 K stage of the DR. After that, it exits the cryogenic environment and is further amplified at room temperature by a Low-Noise Amplifier (LNA). The resulting pulse is combined with a counter-pump tone generated by the RF AWG before entering a downconverter (DNC) with a RF DAC-generated local oscillator LO. The counter-pump tone prevents digitizer (DIG) saturation caused by the pump tone. It has the same frequency as the pump but is phase-shifted to be in counter-phase at the digitizer level. 
Then, the downconverted wave is digitized, split
and IQ-demodulated at signal and idler frequencies taking advantage of hardware acceleration (FPGA).
In principle, the IQ data production process is as follows:
the digitized pulse is multiplied with locally generated digital oscillators, $\text{LO}_{\text{s}} $ and $\text{LO}_{\text{i}}$, for signal and idler tones, respectively. The result is then low-pass filtered to remove high-frequency components, leaving only the baseband pulse, which is subsequently integrated to produce the IQ data for both the signal and idler.
The RF AWG, DNC, and DIG are all part of Keysight's Quantum Control System (QCS). The QCS enables synchronization of the RF DACs in both the RF AWG and DNC, allowing for coherent phase measurements.
\subsection{The Experiment}
\label{The Experiments}
\begin{figure}[htbp] \centerline{\includegraphics[width=0.9\linewidth]{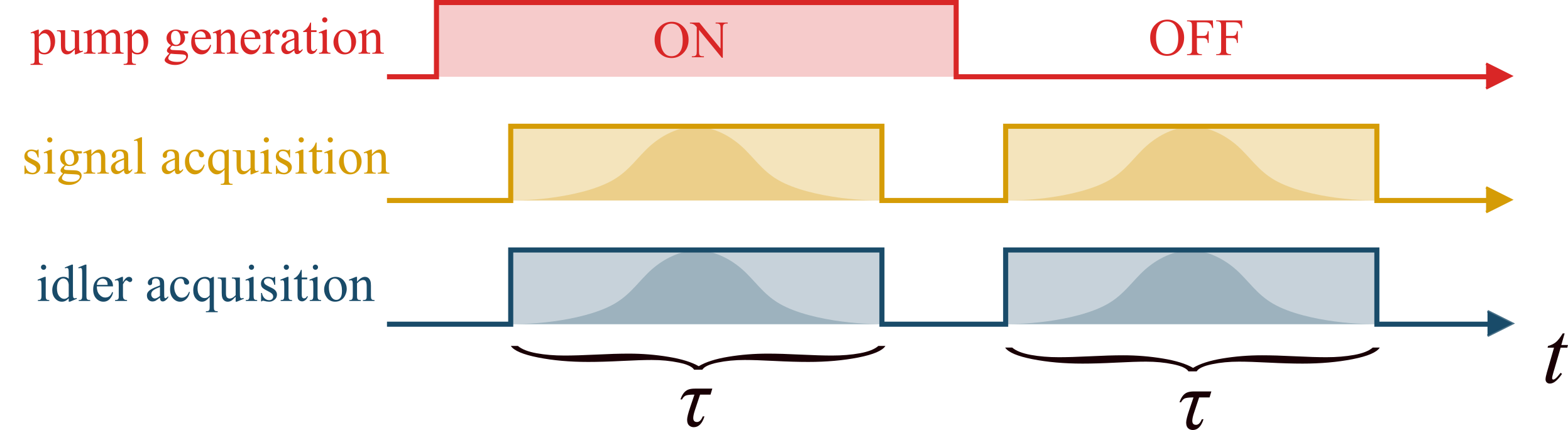}} \caption{Schematic representation of the control and readout pulse sequence for pump on/off generation and measurement at signal and idler frequencies. The acquisition time for both the signal and idler is $\tau$, independent of whether the pump is on or off. The acquisition window shape can be varied. In the presented experiments, rectangular and Gaussian window shapes are explored.}
\label{fig_time} 
\end{figure}

The single acquisition of signal and idler IQ data consists of two stages: one with the pump and counter-pump tones active and one with the tones turned off (see Fig. \ref{fig_time}). 

\begin{figure}[htbp] \centerline{\includegraphics[width=\linewidth]{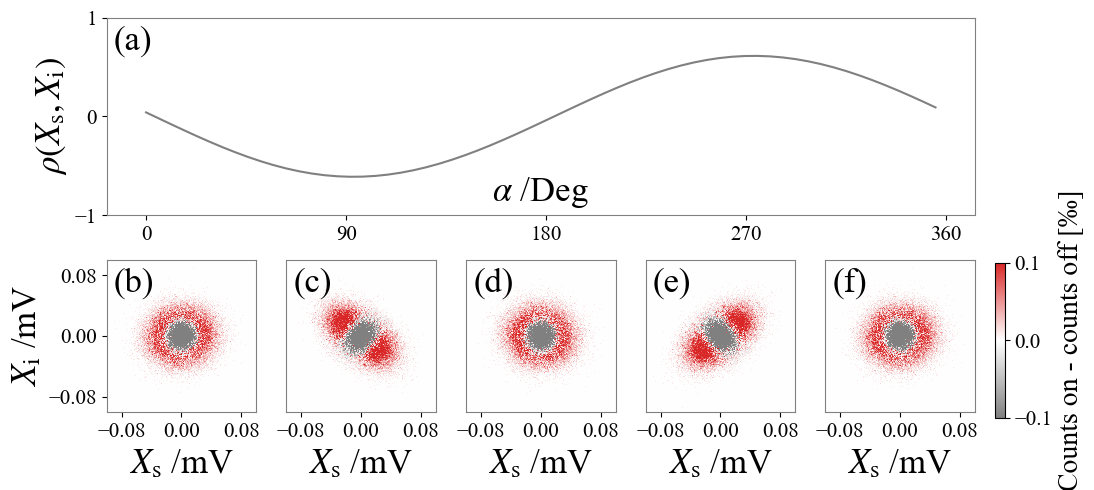}} \caption{In panel (a) the Pearson coefficient $\rho({X_\text{s} X_\text{i}})$ is plotted as a function of the relative phase $\alpha$. In panels (b-f) the histograms of the output field at the digitizer are displayed for $\alpha$ equals to 0°, 90°, 180°, 270°, 360° respectively. $\alpha$ is varied in data analysis by rotating idler's data in the complex plane. Here, the acquisition window has a rectangular shape, and the acquisition time is $\tau=\SI{6}{\micro\second}$.
}
\label{fig_Variance as a function} 
\end{figure}

The latter stage serves as a noise background reference that allows to compute the TMSVS covariance matrix $\sigma^\text{TMSVS}$ and therefore $\rho({X_\text{s}, X_\text{i}})$.
The inferred covariance matrix of the TMSVS is $ \sigma^\text{TMSVS}= \sigma^\text{meas,ON}-\sigma^\text{meas,OFF}+\frac{1}{4}I$, where the last term represents the covariance matrix of the two-mode vacuum state, while $\sigma^\text{meas,ON}$ and $\sigma^\text{meas,OFF}$ are the covariance matrices computed from experimental data with the pump turned on and off, respectively. These matrices are constructed using the signal and idler quadratures, normalized by the gain of the detection chain \cite{Eichler}.
The acquisition time of the two stages is equal and will be denoted as $\tau$. The acquisition window shape can be varied.
The same measurement is repeated for $10^5$ times.
The optimization of correlations (phase matching condition) is achieved by adjusting the relative phase $\alpha = \phi_{\text{LO}_\text{s}}-\phi_{\text{LO}_\text{i}}$, where $\phi_{\text{LO}_\text{s}}$ and $\phi_{\text{LO}_\text{i}}$ are the phases of the digital local oscillator of signal and idler, respectively, within the rotating frame defined by the clock distribution
\cite{Qiu_2023}.
$\rho({X_\text{s}, X_\text{i}})$ as a function of the relative phase $\alpha$, which is varied in data analysis,
is shown in panel (a) of Fig. \ref{fig_Variance as a function}. 
The panels (b-f) in Fig. \ref{fig_Variance as a function} display the histograms of the output field at the digitizer for different $\alpha$ values using a rectangular acquisition time window of $\tau=\SI{6}{\micro\second}$.
This analysis is crucial to target the maximum two-mode correlation and will be used in Section \ref{Characterization of Two-Mode Correlation Linewidths} to assess the two-mode correlation linewidth.

\section{Characterization of Two-Mode Correlation Linewidths}
\label{Characterization of Two-Mode Correlation Linewidths}
This paragraph aims to describe the characterization of the two-mode correlation linewidth of the radiation emitted by the TWPA under test. 
For a given set of pump and idler tones $f_\text p , f_\text i$, the correlation is analyzed around the signal frequency $f_\text s$ given by Eq. \ref{freqmatch}.
With fixed $f_\text{p} = \SI{6.331}{\giga\hertz}$ and idler demodulation frequency $f_\text{i} =\SI{ 6.481}{\giga\hertz}$, the two-mode correlation experiment described in Section \ref{The Experiments} is conducted for 201 values of $f_\text s$ so that $\Delta f$
is in the range between \SI{-1}{\mega\hertz} and \SI{+1}{\mega\hertz}.
The experiment is repeated for four different acquisition times ($3, 4, 5$ and $\SI{6}{\micro\second}$) and with rectangular or Gaussian windowing for the time domain digitalization. The Gaussian window is a Gaussian envelope $E(t)=(1+\alpha)\exp [-2(2t-\tau)^2]-\alpha$ such that $E(0)=E(\tau)=0$ and  $E(\tau/2)=1$.

\begin{figure}[htbp]
\centerline{\includegraphics[width=\linewidth]{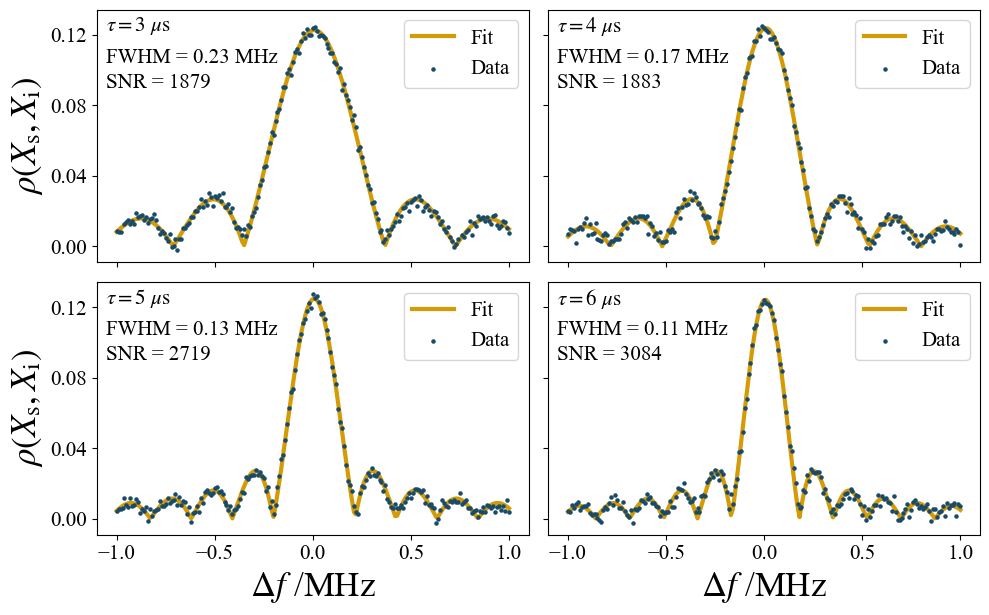}}
\caption{Pearson coefficient $\rho({X_\text{s}, X_\text{i}})$ as a function of the detuning $\Delta f$ for a rectangular window with different acquisition times $\tau$. SNR is computed by rescaling the x-axis to normalize the FWHM across all distributions.
}
\label{fig_rect.}
\end{figure}

\begin{figure}[htbp]
\centerline{\includegraphics[width=\linewidth]{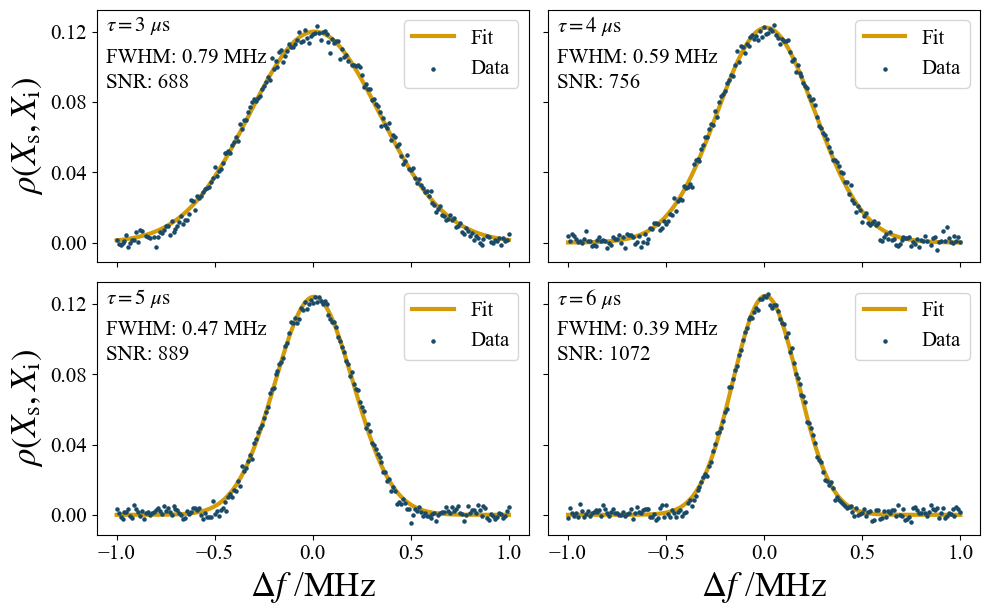}}
\caption{Pearson coefficient $\rho({X_\text{s}, X_\text{i}})$ as a function of the detuning $\Delta f$ for a gaussian window with different acquisition times $\tau$.  SNR is computed by rescaling the x-axis to normalize the FWHM across all distributions.}
\label{fig_gauss.}
\end{figure}

For the rectangular window, the resulting data are fitted in the frequency domain with 
$|A\text{ sinc}(\xi\Delta f)|$ 
(see Fig. \ref{fig_rect.}). 
The use of the absolute value is justified by the fact that for every two-mode correlation experiment the maximum (positive) correlation is searched, as pointed out in Section \ref{The Experiments}. For the Gaussian window, the data are fitted with 
$A \, \text{ exp}({-{(\xi\Delta f)^2}/{2}})$ (see Fig. \ref{fig_gauss.}).
The results show that, in the explored regime, longer acquisition times lead to narrower linewidths, estimated by the Full Width at Half Maximum (FWHM).
Moreover, increasing acquisition time increases the Signal-to-Noise Ratio (SNR).
However, excessively long durations may lead to effects such as frequency drift, potentially counteracting the intended benefits.
Future studies will focus on examining this delicate trade-off.
Regarding the acquisition window, it is clearly seen in Fig. \ref{fig_rect.} and \ref{fig_gauss.} that, for the same acquisition time,
the Pearson coefficient obtained using the rectangular acquisition window displays sharper peaks, reflecting a localized
and highly sensitive response to frequency fluctuations due to its abrupt cutoff. However, this comes at the cost of introducing artifact lobes. In contrast, the
Gaussian window produces a wider peak, while effectively suppressing these lobes.

\section{Conclusions}
\label{Conclusions}
In conclusion, this study emphasizes the critical role of acquisition parameters in quantifying the two-mode correlation linewidth in TWPAs. The Pearson coefficient, which is insensitive to asymmetries in the gain profile, is used to quantify correlations.
The results show that both the shapes of acquisition window and the acquisition duration significantly impact the two-mode correlation linewidth estimation.
Studying the impact of acquisition-induced artifacts on the linewidth is crucial for 
future studies on the intrinsic factors that may influence two-mode correlation linewidth in terms of broadening and frequency shifting. This has relevant applications in quantum illumination (e.g., detecting target velocity) 
and in experimental device characterization (e.g., decoherence time estimation) \cite{Reichert_2022}\cite{PhysRevResearch.3.033095}.

\end{document}